\documentclass[aps,prl,amsmath,showpacs,preprintnumbers,twocolumn,amssymb,groupedaddress,longbibliography]{revtex4-1}

\usepackage{graphicx}
\usepackage{mathptmx, textcomp}
\usepackage[latin1]{inputenc}
\usepackage{tabularx}
\usepackage{siunitx}
\usepackage{color}
\definecolor{bl}{rgb}{0, .1, .6}
\definecolor{rd}{rgb}{1,0,.2}

\usepackage[colorlinks=true, citecolor = bl, linkcolor = bl, urlcolor=bl, pdfborder={0 0 0}]{hyperref}
\usepackage{natbib}
\usepackage{tabularx}
\usepackage{lipsum}
\newcommand{\mum}{\SI{}{\micro\meter}}

\newcommand{\Hz}{{\rm Hz}}
\newcommand{\g}{{\rm G}}
\newcommand{\mg}{{\rm mG}}
\newcommand{\ms}{{\rm ms}}
\newcommand{\be}{\begin{eqnarray}}
\newcommand{\ee}{\end{eqnarray}}
\newcommand{\abg}{a_{\rm bg}}

\newcommand{\add}{a_{\rm dd}}
\newcommand{\edd}{\epsilon_{\rm dd}}

\newcommand{\lc}{\lambda_{\rm c}}
\newcommand{\lp}{\lambda_{\rm p}}

\begin{document}

\title{
Onset of a modulational instability in trapped dipolar Bose-Einstein condensates}

\author{Igor Ferrier-Barbut, Matthias Wenzel, Matthias Schmitt, Fabian Böttcher, Tilman Pfau}
\affiliation{5. Physikalisches Institut and Center for Integrated Quantum Science and Technology IQST,
Universit\"at Stuttgart, Pfaffenwaldring 57, 70550 Stuttgart, Germany}


\begin{abstract}
We explore the phase diagram of a finite-sized dysprosium dipolar Bose-Einstein condensate in a cylindrical harmonic trap. We monitor the final state after the scattering length is lowered from the repulsive BEC regime to the quantum droplet regime. Either an adiabatic transformation between a BEC and a quantum droplet is obtained or, above a critical trap aspect ratio $\lc=1.87(14)$, a modulational instability results in the formation of multiple droplets. This is in full agreement with the predicted structure of the phase diagram with a crossover region below $\lc$ and a multistable region above. Our results provide the missing piece connecting the previously explored regimes resulting in a single or multiple dipolar quantum droplets. 
\end{abstract}

\maketitle

Ultracold dipolar gases constitute a new playground for exploring interaction effects in quantum fluids. In particular because a natural intrinsic length scale emerges, stemming from the combined effects of anisotropic long-range dipole-dipole interaction (DDI) and external forces. This fact, known to the realm of classical magnetic fluids (ferrofluids) for decades \cite{Rosensweig:2013}, has been understood to hold for dipolar Bose-Einstein condensates (dBEC) more recently. The first manifestations were theoretically predicted in finite-size, trapped dBEC: a structuring of the density profile \cite{Goral:2000,Goral:2002,Dutta:2007,Wilson:2008}. In the thermodynamic limit, ref. \cite{Santos:2003} predicted the appearance of a minimum in the dispersion relation of infinite 'pancake' shaped dBECs in the ground state of a harmonic trap. This minimum was called the Roton minimum, inspired by a similar behaviour of the dipersion relation in superfluid helium.\par
These features of the equilibrium many-body state of the quantum fluid were also found to alter its stability. In the infinite case, when changing the parameters (\textit{e.g.}~scattering length $a$ for contact interaction, trapping strength) the Roton minimum softens, and leads to an instability at finite wavelength \cite{Santos:2003}. In finite-size dBECs numerical simulations of the Gross-Pitaevskii equation with DDI predicted the existence of an instability characterized by an ensemble of local collapses \cite{Parker:2009,Wilson:2009}. Such collapse can be understood as the softening of a collective mode ($\omega^2<0$), which belongs to the discrete part of the spectrum where momentum $k$ is not a good quantum number, such that $\eta=\langle k\rangle\times R_{\rm BEC}\lesssim1$ \cite{Ronen:2007}. Here $R_{\rm BEC}$ is the typical size of the BEC and $\langle k\rangle$ is taken for the given mode. The unstable mode differs from the lowest-lying surface and monopole modes. This leads to the appearance of multiple collapses rather than a global collapse. We will refer here both for finite-size ($\eta\lesssim1$) and thermodynamic limit ($\eta\gg1$) to such instabilities as modulational instabilities. In contact-interacting BECs, a modulational instability has been recently reported at negative scattering length \cite{NGuyen:2017,Everitt:2017}. The modulational instability of contact-interacting BECs differs from that of dBECs because all lowest-lying modes are unstable, the favoured wavelength for the instability is set to the mode that has the highest growth rate \cite{NGuyen:2017}. The observation of such modulational instabilities in dBECs was reported for \textsuperscript{164}Dy both in cylindrically symmetric \cite{Kadau:2016} and elongated traps \cite{FerrierBarbut:2016a,FerrierBarbut:2016b}, in the finite-size limit $\eta\lesssim1$. Recent experiments on erbium \cite{Chomaz:2017} report the observation of such instability in the regime $\eta\gg1$ of elongated dBECs, where it can be associated with the softening of a Roton mode.\par
\begin{figure}
\includegraphics[width=\columnwidth]{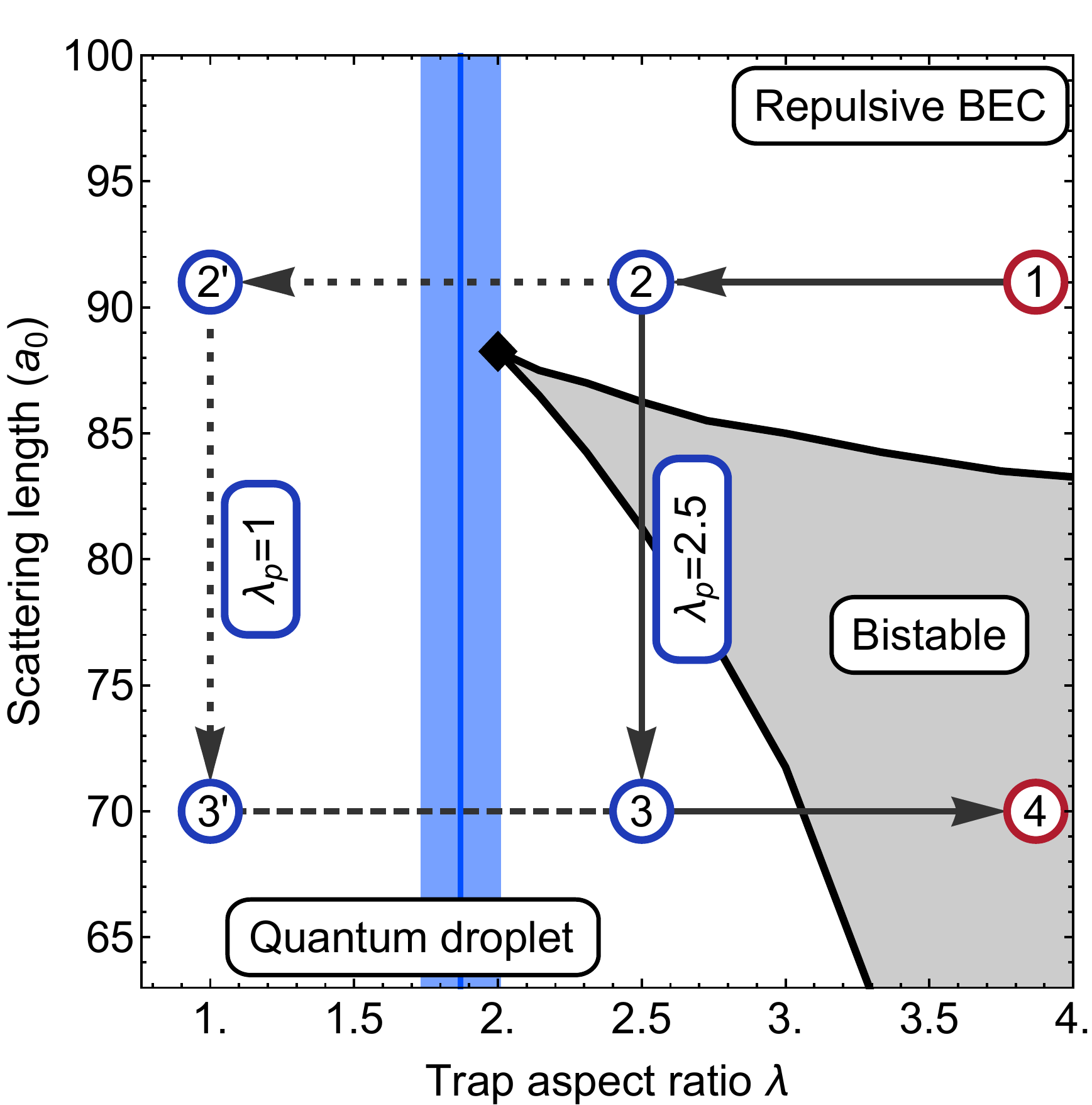}\\
\caption{Phase diagram for a cylindrically trapped \textsuperscript{164}Dy BEC containing 6000 atoms, obtained with our experimental parameters: the $z$ trapping frequency is fixed at $\omega_z=2\pi\times150\,\Hz$. The diagram is calculated with eGPE simulations. For $\lambda$ below the critical point $\lambda<\lc$ shown as a diamond, the repulsive BEC and the quantum droplet states are connected through a crossover. Above the critical point there is a multistable region where both are stable shown (in gray). Lowering the scattering length with $\lambda$ above the critical point leads to a modulational instability. Our experimental procedure to locate the critical aspect ratio $\lc$ is shown as arrows, where we have assumed $\abg=70\,a_0$. We vary $\lp$, two indicative paths are shown [$\lp=2.5$ (through points $2$ and $3$) and $\lp=1.0$ (through points $2'$ and $3'$)], timings are indicated in the text. The vertical blue line and blue area represent the resulting experimental value and error of $\lc=1.87(14)$, see text.}
\label{Fig:phaseDiagram} 
\end{figure}
In the present work, we are interested in the regime $\eta\lesssim1$, of harmonically trapped dBECs in cylindrical traps characterized by the aspect ratio $\lambda=\omega_z/\omega_r$ whee $\omega_{r,z}$ denote trap frequencies and the magnetic field $B$ is oriented along the $z$-axis. We probe this with \textsuperscript{164}Dy. The stability of dBECs as a function of $\lambda$ and $a$ was investigated first with \textsuperscript{52}Cr BECs \cite{Koch:2008}. The dipolar length, defined by $\add=\frac{\mu_0\,\mu^2\,m}{12\pi\hbar^2}$ where $m$ is the atom mass and $\mu$ the magnetic moment, is about eight times smaller for \textsuperscript{52}Cr ($\add=16\,a_0$) than for \textsuperscript{164}Dy ($\add=131\,a_0$). As a consequence, the modulational instability onset occurs for much higher scattering length for Dy than Cr, and takes place even for much smaller samples. No direct evidence for its existence was reported with chromium. In the case of dysprosium, the absence of collapse following the modulational instabilities led to the discovery of quantum droplets stable only within beyond mean-field theory \cite{FerrierBarbut:2016a,Schmitt:2016}, observed also with an erbium dBEC \cite{Chomaz:2016} and in contact interacting BEC mixtures \cite{Petrov:2015,Cabrera:2018,Semeghini:2017}.\par
The discovery of quantum droplets and the realization of the importance of beyond mean-field corrections \cite{FerrierBarbut:2016a} led to new recent works on the stability diagram \cite{Bisset:2016,Wachtler:2016b}. The phase diagram as a function of $\lambda$ and $a$, can be calculated in the framework of the of the extended Gross-Pitaevskii equation (eGPE) which includes an effective term for the beyond mean-field correction, its approximations are discussed in \cite{Wachtler:2016a,Schmitt:2016}. One can then apply a gaussian ansatz to the wave-function which excludes density modulations, or resort to full simulations of the eGPE making no such restriction. The resulting diagram obtained with full simulations is represented in Fig.~\ref{Fig:phaseDiagram}, it contains different regions: At large $a$, a single solution exists, essentially a repulsive BEC with a cloud aspect ratio $\kappa=\sigma_r/\sigma_z$ close to that of the trap, weakly altered by magnetostriction. At very low scattering length, a single solution exists, with an aspect ratio $\kappa\ll1$ largely independent of $\lambda$. This solution corresponds to a quantum droplet that is stabilized by beyond mean-field corrections \cite{FerrierBarbut:2016a,Wachtler:2016b,Bisset:2016}. In between these two regions, and only for trap aspect ratios $\lambda$ larger than a critical value $\lc$, a bistable region exists, where both a repulsive $\kappa\gtrsim1$ mean-field solution and a $\kappa\ll1$ quantum droplet solution are stable. The critical point, shown as a black diamond marks the transition between this bi-stable region and a region where the repulsive BEC and quantum droplets solutions are connected through a crossover. We find in the simulations the region boundaries as in \cite{Wachtler:2016b}, where the two local solutions are found by imaginary time evolution using as starting condition the ground state in the singly-stable large $a$ (small $a$) solution for the $\kappa\gtrsim1$ ($\kappa\ll1$) solution.\par 
Then, the emergence of a modulational instability can be understood from this phase diagram. In the bi-stable region, the fact that a local minimum exists for a mean-field repulsive solution marks the fact that the lowest-lying modes of the system are not soft ($\omega^2>0$) since no global deformation can de-stabilize the gas. Therefore, by lowering $a$ deep into the bistable region, one will observe a modulational instability not taken into account in the Gaussian ansatz, triggered by a higher-lying mode. The authors of \cite{Bisset:2016}, suggested that by following a path in $(\lambda,\,a)$ space that avoids the bi-stable region and lowers $a$ in the crossover region $\lambda<\lc$, one can prepare the quantum droplet ground state. A single quantum droplet was indeed observed by using $\lambda\ll1$ with Er in \cite{Chomaz:2016}. Within the approximations of the extended Gross-Pitaevskii equation, this phase diagram can indeed be directly transposed to other atomic species with different $\add$ like Er, rescaling $a$ and $\bar\omega$ to obtain the same values of $\edd$ and $a/\sqrt{\hbar/m\bar\omega}$.\par

In our experiments we investigate the nature of the instability as scattering length is lowered, in cylindrically trapped \textsuperscript{164}Dy dBECs as described above. We observe the onset of a modulational instability by varying the trap aspect ratio and obtain excellent agreement with theoretical predictions. For this, we implement a cylindrically symmetric harmonic trapping, with frequency $\omega_z$ along the dipoles and $\omega_r$ in the perpendicular plane. Our experiment is described in previous publications \cite{Kadau:2016,Schmitt:2016}. A three-beam optical dipole trap allows for a controllable aspect ratio $\lambda=\omega_z/\omega_r$. Using a Feshbach resonance located at $B_0 = 7.117(3)\,\g$ with width $\Delta B=51(15)\,\mg$, we are able to tune the scattering length above its background value $\abg$ \footnote{The exact value of $\abg$ is still not fully known, with measurements in thermal gases reaching a value of $\abg\simeq90\,a_0$ \cite{Tang:2015,Maier:2015} and critical number measurement in self-bound droplets implying $\abg\simeq63\,a_0$ \cite{Schmitt:2016}. For the purpose of the present work, the exact $\abg$ is not directly relevant as we will develop later.}.\nocite{Tang:2015,Maier:2015}\par
\begin{figure}[hbtp]
\includegraphics[width=\columnwidth]{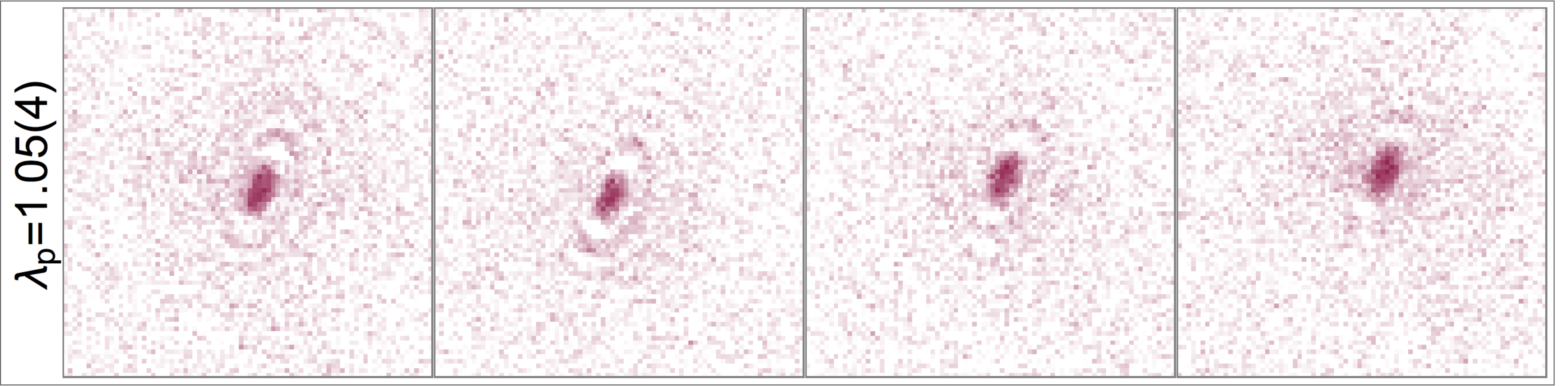}\\
\includegraphics[width=\columnwidth]{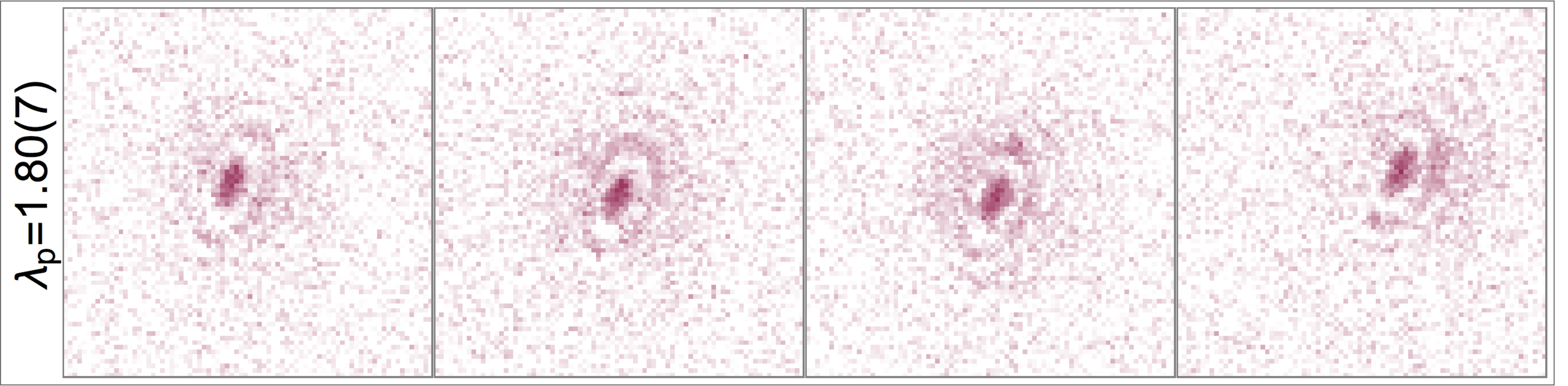}\\
\includegraphics[width=\columnwidth]{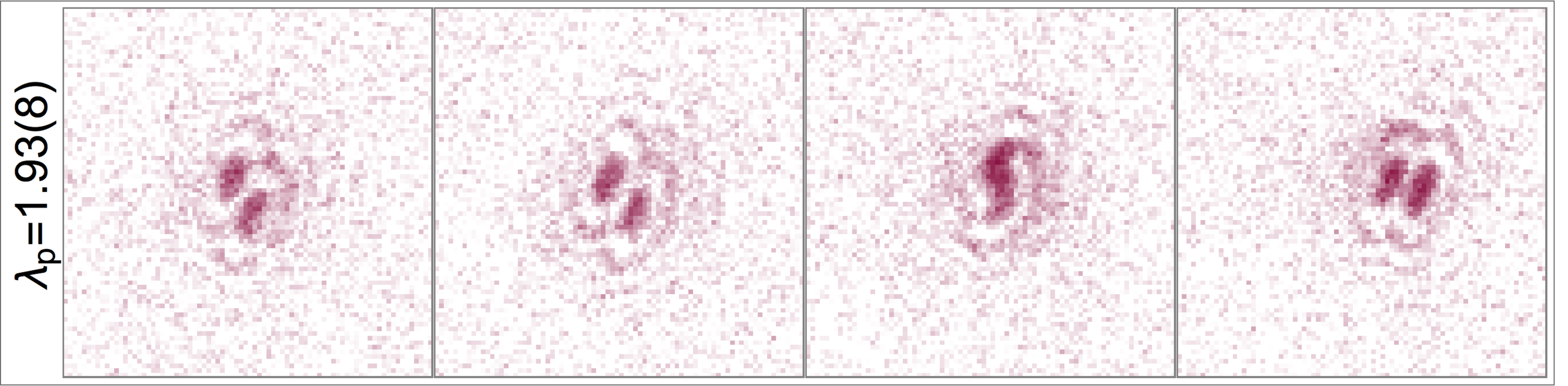}\\
\includegraphics[width=\columnwidth]{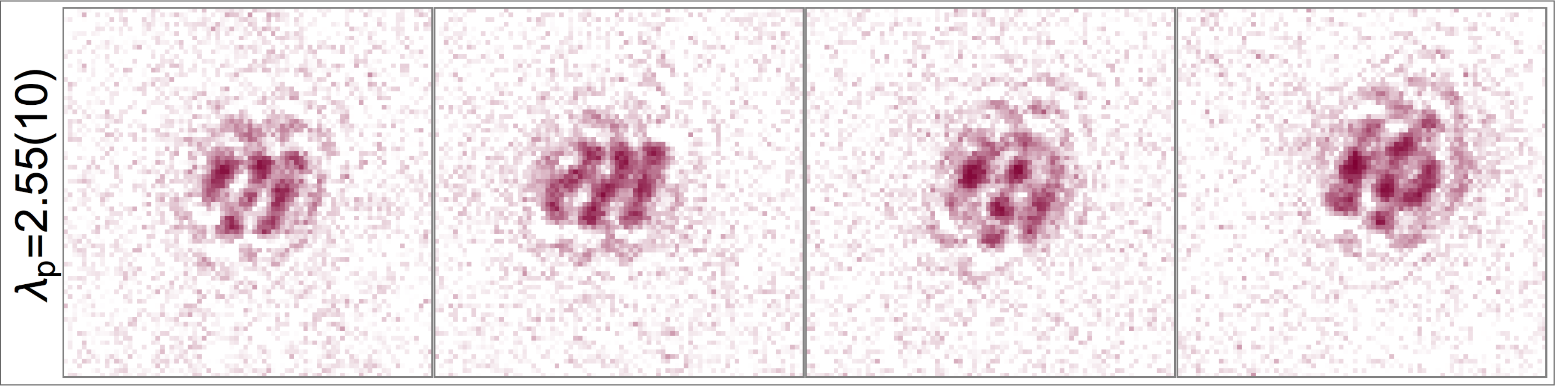}\\
\caption{Examples of final integrated density images resulting from different paths (identical within each row), indicated in the row. Imaging fringes are visible due to the size of the quantum droplet being smaller than our imaging resolution, and likely a slight displacement out of the focal plane. Misalignment of the objective caused a distortion of the images. The field of view in each image is $26\,\mum\times26\,\mum$.}
\label{Fig:fourRows}
\end{figure}

Here, we restrict our exploration to aspect ratios $\lambda<3$. For low enough $a$, the groundstate of the system is a single quantum droplet \footnote{In another geometry with very confining traps, this is no longer the case \cite{Wenzel:2017}.}\nocite{Wenzel:2017}. However, depending on the path taken in phase space $\{\lambda,\,a\}$, we observe that the final state differs. If crossing the instability ($\lambda>\lc$) one obtains an excited state, comprised of several droplets. The final density distribution is shaped by post-instability non-linear dynamics and does not allow to study in depth the instability itself for instance identifying the unstable mode triggering the instability. But the observation of multiple droplets indicates that a modulational instability was crossed. If lowering $a$ at $\lambda<\lc$ in the crossover part of the phase diagram, then a single droplet should be obtained. To locate the onset of the modulational instability, we apply the following procedure, which is closely related to what is suggested in \cite{Bisset:2016}: All our experiments start with a Bose-Einstein condensate of $\approx6000$ \textsuperscript{164}Dy atoms, in a $\lambda_0=3.87(5)$ trap. The $z$ trap frequency is kept constant $\omega_z=2\pi\times150(3)\,\Hz$ while the radial one $\omega_r$ is varied to control $\lambda$. The initial scattering length is $a=1.3(2)\,\abg$ (at $B=6.968(5)\,\g$), point 1 in Fig.~\ref{Fig:phaseDiagram}. The final point is fixed (point 4), with the same aspect ratio $\lambda_0$, but at the background scattering length within experimental uncertainty $a=1.00(5)\,\abg$ ($B=5.468(5)\,\g$). The path from the initial to final conditions is taken as follows, shown in Fig.~\ref{Fig:phaseDiagram}: First (point 1 to 2), the trap aspect ratio is lowered in $50\,\ms$ to an arbitrary value $1.05<\lp\leqslant2.55$. Second (2 to 3) the scattering length is lowered in $20\,\ms$ to the final value, third the aspect ratio is brought back in $50\,\ms$ to $\lambda_0$ (3 to 4). The only variable describing the path taken is thus $\lp$. The mean trapping frequency varies between $\bar\omega=2\pi\times59\,\Hz$ ($\lp=2.55$), $\bar\omega=2\pi\times143\,\Hz$ ($\lp=1.05$). The cloud is then immediately imaged in situ using phase-contrast imaging.\par
The final density distribution, shaped by beyond mean-field effects, retains a strong memory of the path taken as visible in Fig.~\ref{Fig:fourRows}. For low aspect ratios, we observe a single droplet with little shot-to-shot variability, confirming expectations. This is independent of aspect ratio up to a threshold. On the other hand, for large $\lp$ the resulting density distribution is excited with several droplets that do not decay to the single droplet groundstate, and the observed density shows a strong shot-to-shot variability.\par
 To quantify this behaviour, we apply a fit-free image analysis to the integrated density images. It is based on the Principal Component Analysis (PCA) method \cite{Jolliffe:1986}. PCA has been used to analyze experiments in ultracold atoms in the past, see for instance \cite{Segal:2010,Dubessy:2014} and references therein. We apply it to our entire image data set $\{I_{k,s}\}$, $(k\in [1,\,N]$, $s\in[1,\,P])$. The set contains $N=709$ images, each composed of $P=80\times80$ pixels (chosen to give reasonable computational time). For this, we construct the correlation matrix $C_{mn}=\frac1N\sum_{k=1}^N\delta I_{k,m}\delta I_{k,n}$ where $\delta I_k=I_k-\langle I\rangle$ and $\langle I\rangle$ is the mean of all images. The individual images are then decomposed on the basis $\{v_l\}$ of eigenvectors of $C$: $\delta I_k=\sum_{l=1}^{N-1}\alpha_{l,k}v_l$ \footnote{Naively, since the dimension of $C$ is $P\times P$, the sum would extend to $l=P$, but in fact given the definition of $C$, its rank is only $N$. The first component of the PCA decomposition is the dataset mean, which is here zero by definition of $\delta I_k$. Thus the sum can be stopped at $l=N-1$}. The components $l$ are sorted by decreasing eigenvalue $\eta_l$ of $C$. This decomposition allows a quantitative comparison of the images. It holds a clear signature of the path taken, and provides a precise way to measure the critical aspect ratio $\lc$.\par
\begin{figure}[hbtp]
\hspace{1cm}\includegraphics[width=.85\columnwidth]{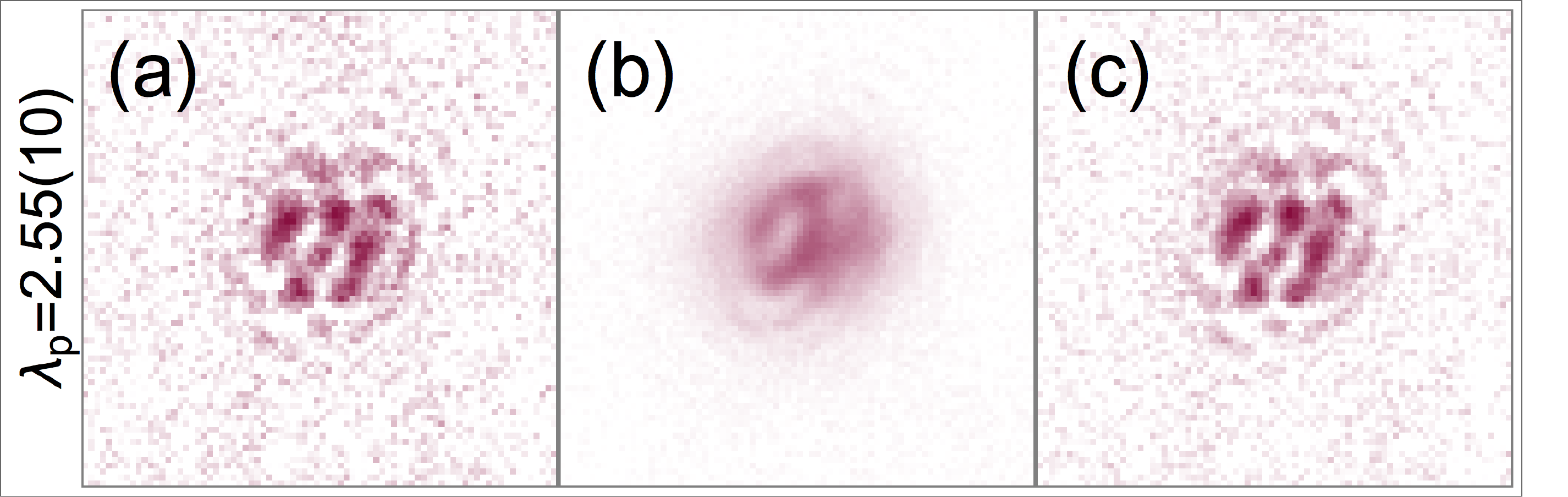}\\
\hspace{1cm}\includegraphics[width=.85\columnwidth]{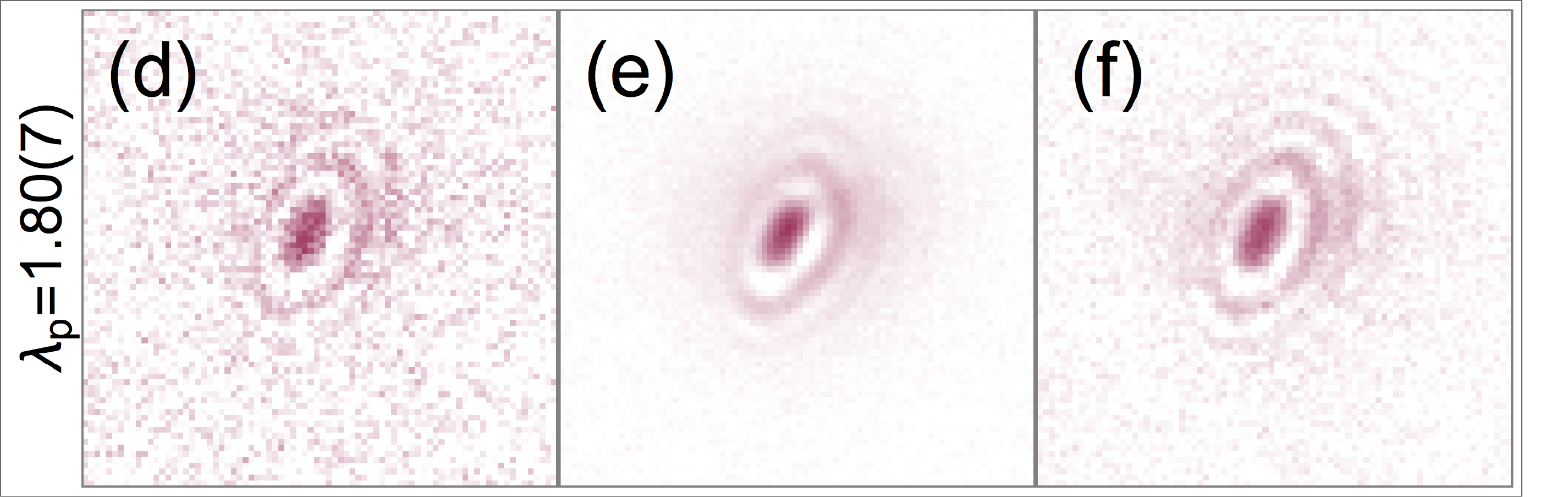}
\includegraphics[width=\columnwidth]{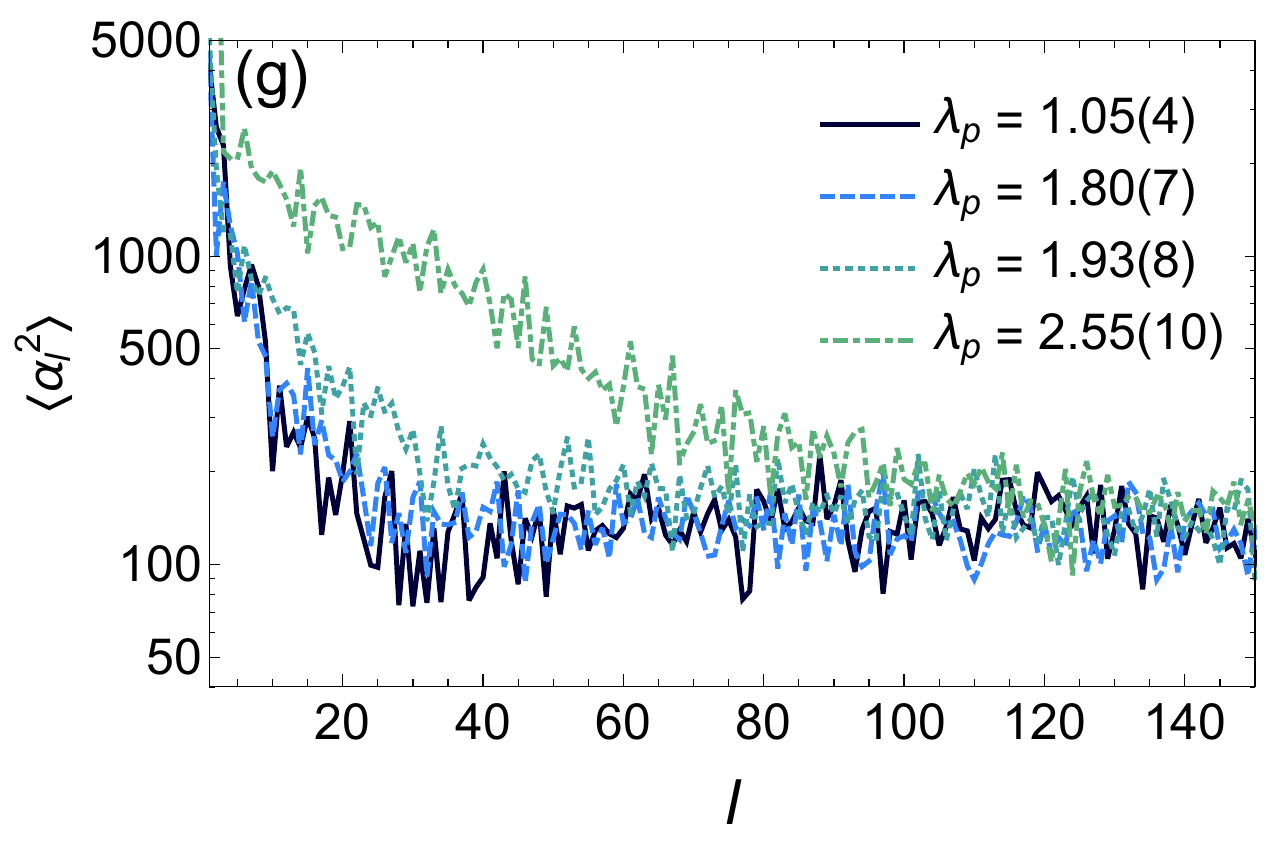}\\
\caption{(a)-(f) PCA image decomposition examples, comparing two different paths: (a)-(c) for $\lp=2.55(10)$ and (d)-(f) for $\lp=1.80(7)$.  (a) and (d): raw images, (b) and (e): decomposition of the raw image only on the first 5 eigenvectors. The result for $\lp=1.80$ is very close to the raw image, while for $\lp=2.55$ it is markedly different. (c) and (f) decomposition over 100 eigenvectors. The field of view in each image is $26\,\mum\times26\,\mum$. (g) Comparing the average distribution of squared eigenvalues $\alpha_l^2$ for four different paths. Below $\lp\approx1.9$ they are indistinguishable from each other while above that value a clear difference is observed with more eigenvectors contributing.}
\label{Fig:alphadecays}
\end{figure}

Due to weak shot-to-shot fluctuations, in particular in the center-of-mass visible in Fig.~\ref{Fig:fourRows}, each image $k$ has a different $\{\alpha_{l}\}$ decomposition. Thus we plot the average distribution of $\langle\alpha_l^2\rangle$ over all images taken after the same path. One indeed observes that it decays very differently depending on the path $\lp$ as can be seen in Fig.~\ref{Fig:alphadecays}. For low aspect ratios $\lp$, the average distribution of eigenvalues $\langle\alpha_l^2\rangle$ decays fast. Furthermore, the different distributions for different $\lp$ are indistinguishable from each other. On the other hand, for large aspect ratios, this distribution decays much more slowly, and varies with $\lp$.\par
\begin{figure}[hbtp]
\includegraphics[width=\columnwidth]{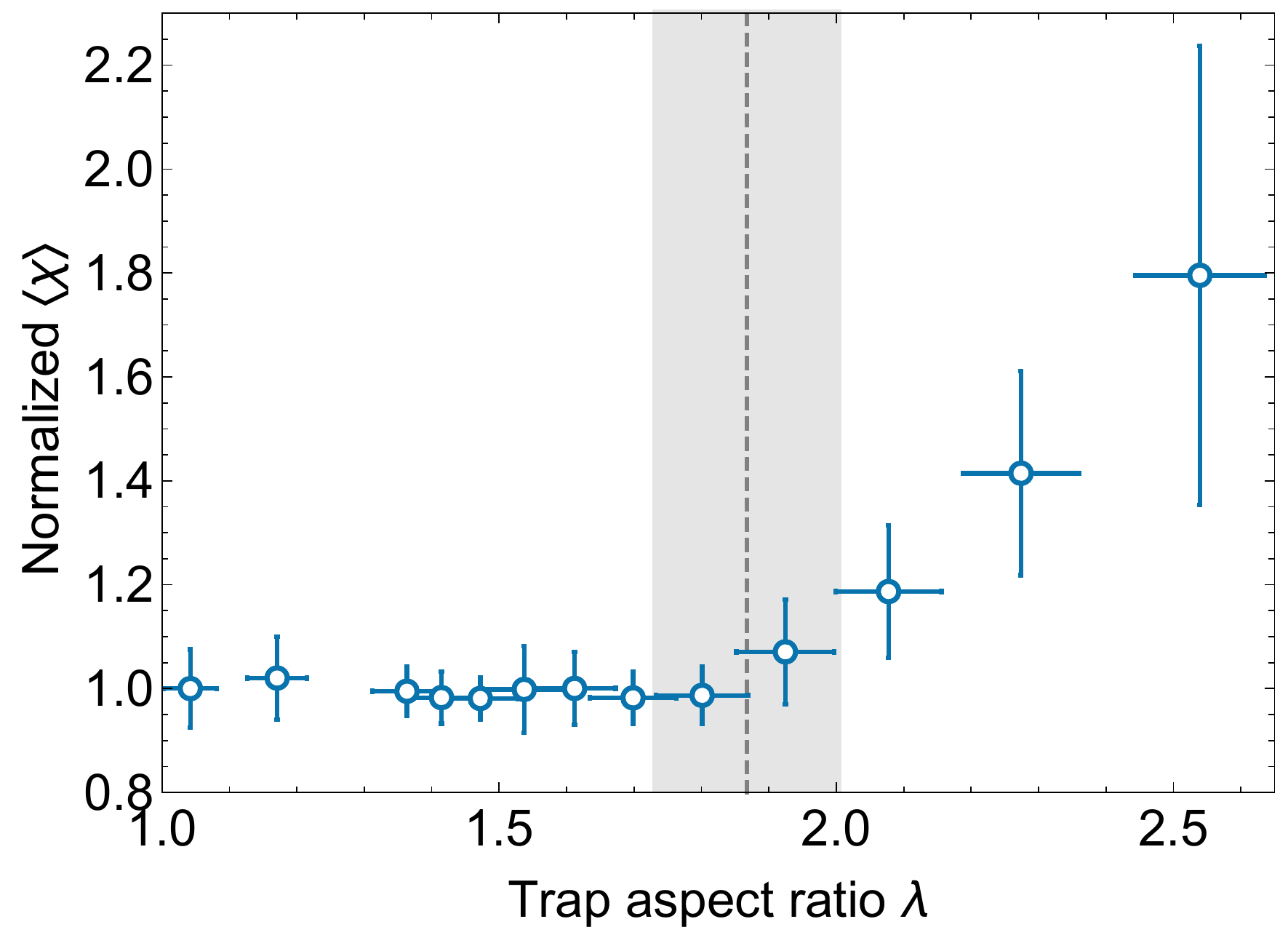}\\
\caption{$\langle\chi\rangle$ as a function of $\lp$. The average is obtained from over $50$ images for each $\lp$, and the values are normalized to the lowest $\lp$. The standard deviation of $\chi_k$ is shown as a vertical error bar. The gray area represents the confidence interval of the extracted $\lc$.}
\label{Fig:Complexity}
\end{figure}

Many observables using PCA can allow to differentiate the images, we find that they all lead to the same conclusion, and we choose to calculate the quantity $\chi_k$ for image $I_k$ defined as: 
\begin{equation}
\chi_k=\sum_{l=1}^{N-1}(\alpha_{l,k})^2=\sum_{s=1}^{P}(\delta I_{k,s})^2
\end{equation}
The last identity shows that PCA decomposition is in fact not even necessary to obtain $\chi_k$ and that only a straightforward analysis can be done, simply taking the square of the image after having subtracted the mean of all images. Therefore this method measures how much each image differs from the overall mean, $\chi_k$ is of course increased when a large shot-to-shot variability occurs. Our analysis is thus similar to what was implemented for instance in \cite{Chen:2011,Meldgin:2016}. However the PCA decomposition was necessary to show that the $\langle\alpha_l^2\rangle$ do not differ from each other, and that one can indeed use only $\chi_k$. In the PCA decomposition $\chi_k$ also reflects the `complexity' of the image $I_k$ in the sense that the more eigenvectors need to be added and subtracted to reproduce the image, the higher $\chi_k$ is. We then calculate the average $\langle\chi\rangle$ again for all images obtained from the same path $\lp$. The absolute value of $\langle\chi\rangle$ is arbitrary, it depends for instance on the number of atoms. Therefore we normalize it to the value obtained for the lowest $\lp=1.05$. The resulting normalized $\langle\chi\rangle$ as a function of $\lp$ is represented in Fig.~\ref{Fig:Complexity}. It exhibits a very clear threshold behaviour, with unchanged $\langle\chi\rangle$ up to a critical value $\lc$ followed by a linear increase above this threshold. The fact that below threshold, the images do not differ significantly from each other, while above threshold they vary strongly is thereby quantified. Then from the data of Fig.~\ref{Fig:Complexity} the critical aspect ratio can be extracted, we observe that $\langle\chi\rangle$ departs from its background value at:
\begin{equation}
	\lc=1.87(14),
\end{equation}
with the confidence interval (gray area) given by the experimental uncertainty on $\lambda$. The exact $\lc$ value, represented as a blue area in Fig.~\ref{Fig:phaseDiagram} is in slightly better agreement with predictions of the extended Gross-Pitaevskii equation with our experimental parameters ($\lc=2.0$, black diamond in Fig~\ref{Fig:phaseDiagram}) than expected using a Gaussian ansatz ($\lc=1.6$, not shown).
\par
To conclude, our experiments start in a region where the repulsive Bose-Einstein condensate is stable. The fact that we are able to adiabatically prepare quantum droplets implies that when lowering $\lambda$ at fixed $a$ no modulational instabilty is crossed. This means that the initial conditions are most likely in the single-solution repulsive BEC region as shown in Fig.~\ref{Fig:phaseDiagram}. Given our uncertainty on the Feshbach resonance used to modulate the scattering length our uncertainty in $a$ is about $20\,\%$ in units of $\abg$. In consequence we obtain a weak lower bound on $\abg\geqslant58\,a_0$ in agreement with all expectations. Our final scattering length value is within the multistable region where within the Gaussian ansatz a repulsive BEC could be stable. Due to a modulational instability not taken into account within this theory, the BEC is unstable and forms multiple droplets. Thus, we have measured the experimental critical aspect ratio for the modulational instability of a dipolar Bose-Einstein condensate by a very straightforward data analysis involving no fit to the data. This provides the missing piece between the regimes explored in \cite{Kadau:2016} ($\lambda\simeq3$) and in \cite{Chomaz:2016} ($\lambda\simeq0.1$), thus being a powerful confirmation of the structure of the phase diagram.\par
 Locating this critical point is an important benchmark for all studies of harmonically trapped dipolar Bose-Einstein condensates, it allowed us to produce single self-bound droplets \cite{Wachtler:2016a,Baillie:2016} which we reported in \cite{Schmitt:2016}. The natural continuation of our work is to locate the instability line as a function of $\lambda$ and $a$ above $\lc$. This line cannot be theoretically predicted within the Gaussian ansatz. Experimentally, it will depend on other parameters such as temperature since thermal fluctuations might trigger the instability. Another direction is to understand the relationship between the modulational instabilities studied here and the soliton fragmentation reported for BEC mixtures in \cite{Cheiney:2017}.\par
 
\begin{acknowledgments}

The authors thank Tim Langen for critical reading of the manuscript, and Luis Santos for discussions.  This work is supported by the German Research Foundation (DFG) within FOR2247. IFB acknowledges support from the EU within a Horizon2020 Marie Sk\l odowska Curie IF (703419 DipInQuantum).
\end{acknowledgments}

\bibliography{CriticalPoint}

\end{document}